\documentclass[runningheads]{llncs}
\usepackage{graphicx}
\usepackage{xcolor}
\usepackage{booktabs}
\usepackage{nicefrac}
\usepackage{hyperref}
\usepackage{marvosym}
\graphicspath{{pics/}}

\newcommand{\cmmnt}[1]{}

\begin{document}

\title{Assessing Reliability and Challenges of Uncertainty Estimations for Medical Image Segmentation}
\titlerunning{Reliability and Challenges of Uncertainty Estimations}
%

\author{Alain Jungo\inst{1,2}\textsuperscript{(\Letter)}\orcidID{0000-0001-8327-4653} \and Mauricio Reyes\inst{1,2}}

\authorrunning{A. Jungo and M. Reyes}
%
\institute{Insel Data Science Center, Inselspital, Bern University Hospital, University of Bern, Bern, Switzerland \\
ARTORG Center, University of Bern, Bern, Switzerland\\
\email{alain.jungo@artorg.unibe.ch}}
\maketitle              
\begin{abstract}
Despite the recent improvements in overall accuracy, deep learning systems still exhibit low levels of robustness. Detecting possible failures is critical for a successful clinical integration of these systems, where each data point corresponds to an individual patient. Uncertainty measures are a promising direction to improve failure detection since they provide a measure of a system's confidence. Although many uncertainty estimation methods have been proposed for deep learning, little is known on their benefits and current challenges for medical image segmentation. Therefore, we report results of evaluating common voxel-wise uncertainty measures with respect to their reliability, and limitations on two medical image segmentation datasets. Results show that current uncertainty methods perform similarly and although they are well-calibrated at the dataset level, they tend to be miscalibrated at subject-level. Therefore, the reliability of uncertainty estimates is compromised, highlighting the importance of developing subject-wise uncertainty estimations. Additionally, among the benchmarked methods, we found auxiliary networks to be a valid alternative to common uncertainty methods since they can be applied to any previously trained segmentation model. 

\end{abstract}
\section{Introduction}

Deep learning-based methods have led to impressive improvements in medical image segmentation over the past years. For many tasks, the performance is comparable to human-level performance, or even surpasses it \cite{litjens2017survey}. Nonetheless, despite improvements in accuracy, the robustness aspects of these systems call for significant improvements for a successful clinical integration of these technologies, where each data point corresponds to an individual patient. This highlights the importance of having mechanisms to effectively monitor computer results in order to detect and react on system's failures at the patient level. Among others, uncertainty measures are a promising direction since uncertainties can provide information as to how confident the system was on performing a given task on a given patient. This information in turn can be used to leverage the decision-making process of a user, as well as to enable time-effective corrections of computer results by for instance, focusing on areas of high uncertainty.

Different approaches have been proposed to quantify uncertainties in deep learning models. Among the most popular approaches are: a) Bayesian uncertainty estimation via test-time dropout \cite{gal2016dropout}, b) aleatoric uncertainty estimation via a second network output \cite{kendall2017uncertainties}, and c) uncertainty estimation via ensembling of networks \cite{lakshminarayanan2017simple}. In medical image segmentation, uncertainty measures are of interest at three levels. The first, most fine-grained level, is the voxel\footnote{For simplicity, we use \textit{voxel} even if it could be a two-dimensional image.}-wise uncertainty, which provides a measure of uncertainty for the predicted class of each voxel. This level of uncertainty is especially useful for the interaction with humans, be it by providing additional information to foster comprehensibility or as guidance for correction tasks. The second level is the uncertainty at the level of a segmented instance (or object). Nair et al. \cite{nair2018exploring} and Graham et al. \cite{graham2019mild} used instance-level uncertainty to reduce the false discovery rate of brain lesions and cells, respectively. In both approaches voxel-wise uncertainties were aggregated to obtain an instance-wise uncertainty. Similarly, Eaton-Rosen et al. \cite{eaton2018towards} aggregated voxel-wise uncertainties of brain tumor segmentations to obtain confidence intervals for tumor volumes. The third level is the subject-level uncertainty, which informs us whether the segmentation task was successful (e.g., above a certain metric). Having information about success or failure would be sufficient for many tasks, e.g., high-throughput analysis or selection of cases for expert review. As proposed by Jungo et al. \cite{jungo2018uncertainty}, task-specific aggregation of the voxel-wise uncertainties could be used to obtain subject-level uncertainties. In contrast, DeVries et al. \cite{devries2018leveraging} and Robinson et al. \cite{robinson2018real}, proposed an auxiliary neural network that predicts segmentation performance at the subject-level. A current challenge to use these latter type of approaches is that considerable large training datasets are necessary in practice to ensure their reliability \cite{devries2018leveraging}. 

In order to better understand the benefits and current challenges in uncertainty estimation for medical image segmentation, we evaluated common uncertainty measures with respect to their reliability, their benefit, and limitations. Additionally, we analyzed the requirements for uncertainties in medical image segmentation and we make practical recommendations for their evaluation.

\section[Material \& Methods]{Material \& Methods\footnote{Code available at \url{https://github.com/alainjungo/reliability-challenges-uncertainty}}}
\subsection{Data}
We selected two publicly available, and distinct datasets for the experiments. The first dataset is the brain tumor segmentation (BraTS) challenge dataset 2018 \cite{bakas2018identifying} consisting of 265 subjects. Each subject features four magnetic resonance images (T1-weighted, T1-weighted post-contrast, T2-weighted, FLAIR) of a size of 240$\times$\allowbreak240$\times$\allowbreak155 isotropic (1 mm\textsuperscript{3}) voxels. We split the dataset into 100 training, 25 validation, and 160 testing subjects, combined the three tumor sub-compartment labels to segment the whole tumor, and performed a z-score intensity normalization ($\mu=0, \sigma=1$) on each subject and image individually. The second dataset is the international skin imaging collaboration (ISIC) lesion segmentation dataset 2017 \cite{codella2018skin} consisting of 2000 training, 150 validation, and 600 testing images. We resized the color images to a size of 256$\times$192 pixels and normalized the intensities to the range $[0,1]$. 

\subsection{Experimental setting}
\label{sec:architecture}
Our aim is to evaluate the reliability of uncertainty measures for deep learning-based segmentation of medical images. Rather than building a specific fine-tuned, top-performing segmentation model, we used a U-Net-like architecture \cite{ronneberger2015u} due to its popularity, simplicity, and to minimize architectural influences on the outcomes\footnote{We also conducted experiments with a DenseNet-like architecture with no notable differences in the outcome and therefore omit it here for space and clarity reasons.}. The architecture consists of four pooling/upsampling steps and has dropout regularization (p=0.05) and batch normalization after each convolution. We used a common training scheme consisting of a cross-entropy loss with Adam optimizer (learning rate: $10^{-4}$), and applied early stopping with respect to the validation set Dice coefficient. Any adaptation to this architecture and training scheme was performed to fit the needs of each studied uncertainty approach. 

\subsection{Uncertainty methods}
\label{sec:measures}
We evaluated the following five different uncertainty measures:
\\
\textbf{Baseline uncertainty: Softmax entropy.}
Although the softmax output of a model is arguably a probability measure \cite{gal2016dropout}, we considered it as reference comparison as it is implicitly generated by segmentation networks. We named this strategy \textit{baseline}. We used the normalized entropy $H=\nicefrac{-\sum_{c \in \mathcal{C}}{p_c log \left(p_c\right)}}{log(|\mathcal{C}|)} \in [0, 1]$ as a measure of uncertainty, where $p_c$ is the softmax output for class $c$ and $\mathcal{C}$ is the set of classes ($\mathcal{C} = \{0,1\}$ in our case).  
\\
\textbf{MC dropout.}
Test time dropout can be viewed as an approximation of a Bayesian neural network \cite{gal2016dropout}. $T$ stochastic network samples can be interpreted as Monte-Carlo samples of the posterior distribution of the network's weights and result in a class probability of $p_c = \nicefrac{1}{T}\sum_{t=1}^{T}{p_{t,c}}$. We employed the normalized entropy of these probabilities as a measure of uncertainty. For the experiments, we used $T=20$ and considered two different dropout layer positioning strategies. First, we applied MC dropout on the base model (see Sec.~\ref{sec:architecture}), which uses minimal dropout (p=0.05) after each convolution. Second, we applied more prominent dropout (p=0.5) at the center positions (i.e., before pooling and after upsampling, similar to \cite{nair2018exploring}). Accordingly, we name these two strategies as \textit{baseline+MC} and \textit{center+MC}. 
\\
\textbf{Aleatoric uncertainty.}
In contrast to the model uncertainty (captured by e.g. MC dropout), the aleatoric uncertainty is said to capture noise inherent in the observation \cite{kendall2017uncertainties}. It is obtained by defining a network $f$ with two outputs $[\hat{x}, \sigma^2] = f(x)$ and input $x$, where the outputs $\hat{x}$ and $\sigma^2$ are the mean and variance of the logits perturbed with Gaussian noise. The aleatoric loss optimizes both outputs simultaneously by MC sampling (ten samples in our case) of the perturbed logits. We used $\hat{x}$ for the class predictions and the variance $\sigma^2$ as a measure of uncertainty. We normalized the variance to $[0,1]$ over all predictions.
\\
\textbf{Ensembles.}
Another way of quantifying uncertainties is by ensembling multiple models \cite{lakshminarayanan2017simple}. We combined the class probabilities of each network $k$ by the average $p_c=\nicefrac{1}{K}\sum_{k=1}^{K}p_{k,c}$ over all $K=10$ networks and used the normalized entropy as uncertainty measure. The individual networks share the same architecture (see Sec.~\ref{sec:architecture}) but were trained on different subsets (90\%) of the training dataset and different random initialization to enforce variability.  
\\
\textbf{Auxiliary network.}
Inspired by \cite{devries2018leveraging,robinson2018real}, where an auxiliary network is used to predict segmentation performance at the subject-level, we apply an auxiliary network to predict voxel-wise uncertainties of the segmentation model by learning from the segmentation errors (i.e., false positives and false negatives). For the experiments, we considered two opposing types of auxiliary networks. The first one, named \textit{auxiliary feat.}, consists of three consecutive 1$\times$1 convolution layers cascaded after the last feature maps of the segmentation network. The second auxiliary network, named \textit{auxiliary segm.}, is a completely independent network (same U-Net as described in Sec.~\ref{sec:architecture}) that uses as input the original images and the segmentation masks produced by the segmentation model (generated by five-fold cross-validation). We normalized the output uncertainty subject-wise to $[0,1]$ for comparability purposes.

\subsection{Assessing quality of uncertainties}
We adopted three metrics to evaluate the quality of uncertainties. Additionally, we computed the Dice coefficient to also verify segmentation performance as uncertainty methods typically link both tasks.
\\
\textbf{Calibration.}
Model calibration is important when not only the predicted class but also its corresponding confidence is of interest. In this regards, calibration has been used as a surrogate to asses the reliability of uncertainties \cite{kendall2017uncertainties}. A model is said to be perfectly calibrated if its predictions $f(x)$ with confidence $p$ do occur with a fraction $p$ of the time ($P(y=1|f(x)=p)=p$ for the binary case). Meaning for example that for 100 predictions with a confidence of 0.7, 70 predictions are expected to be correct \cite{guo2017calibration}. We assessed calibration of uncertainties by reliability diagrams and expected calibration error (ECE) \cite{guo2017calibration}. Reliability diagrams show the deviation of the perfect calibration by plotting the binned predicted confidences against the accuracy obtained for each bin (fraction of positives). The ECE is defined as the absolute error of these bins (i.e., the gap between confidence and accuracy) weighted by the number of samples in the bins, where a lower ECE (close to zero) indicates a better calibration. In our experiments, we used a bin size of ten and used the model output probabilities as confidence. For methods not providing segmentation probabilities but direct segmentation uncertainty estimates (i.e., auxiliary and aleatoric), we translated the uncertainties by $y(1 - 0.5q) + (1 - y)(0.5q)$ to confidences, where $y\in\{0,1\}$ is the segmentation label and $q\in[0,1]$ is the normalized uncertainty. 
\\
\textbf{Uncertainty-Error overlap.}
In a practical setting, perfect calibration of a model is impossible \cite{guo2017calibration}. Often, segmentation tasks do not require perfect calibration but it would be sufficient for a model to be uncertain where it makes mistakes and certain where it is correct. To assess this condition, we used the overlap (determined by the Dice coefficient) between the segmentation error and the thresholded uncertainty, termed \textit{uncertainty-error overlap} (\textit{U-E}). This metric is not influenced by the true negatives from background areas, which are typically enormous in medical image segmentation. It is therefore an alternative for the ECE, which includes foreground as well as background areas.
\\
\textbf{Corrections.}
Motivated by previous works using uncertainty estimations, we assessed the quality of uncertainties by evaluating their benefit to correct segmentations. We define TPU, TNU, FPU, FNU as uncertainty in the true positives (TP), true negatives (TP), false positives (FP), and false negatives (FN). A beneficial correction is said to improve the Dice coefficient, hence, to benefit from removal of false positives, the relation $FPU\:(TP) > TPU\:(TP + FP + FN)$ needs to be satisfied (for the accuracy $FPU> TPU$ is sufficient). Similarly, in order to benefit from adding voxels (i.e., correct false negatives), the relation $FNU\:(TP + FP + FN)>TNU\:(TP)$, needs to be satisfied. However, the latter relation is not practically applicable due to large backgrounds and thus typically large $TNU$. Since voxel-wise corrections (as opposed to instance-wise corrections) might be more harmful than beneficial, we calculated the proportion of subjects that fulfill the benefit condition for false positive removal, \textit{BnF}, as means of comparison to other methods.

\section{Results}

\begin{figure}[t]
\includegraphics[width=\textwidth]{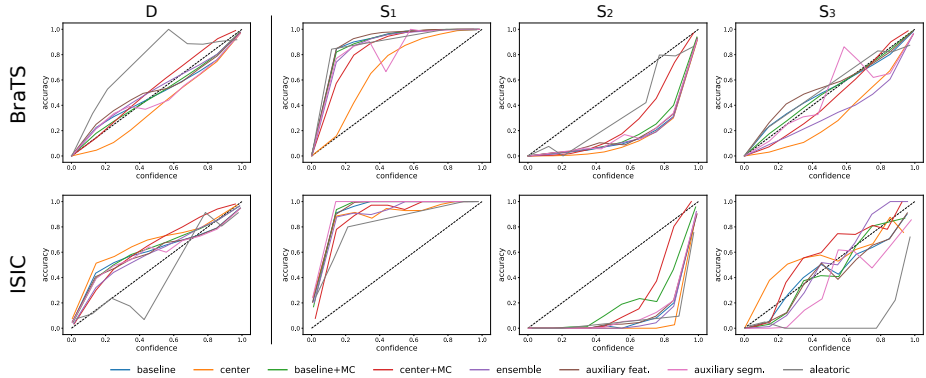}
\caption{Calibration at the dataset level (D) compared to the (mis)calibration at the subject level (S1, S2, S3) for the different uncertainty methods. S1, S2, S3 correspond to exemplary subjects for which the models are underconfident (S1), overconfident (S2), and well-calibrated (S3). Rows correspond to results on the BRATS and ISIC datasets.} 
\label{fig:calibration}
\end{figure}

Fig.~\ref{fig:calibration} compares the calibration at the dataset level (i.e., all voxels in the dataset) with the calibration at the subject level (i.e., voxels of one subject). It shows the miscalibration that can occur at subject level (S1 and S2) while the calibration at dataset level is good. We found approximately 28\%/46\% underconfident and 32\%/18\% overconfident calibrations for the subjects of the BraTS/ISIC dataset. This underlines the special caution needed when using the calibration-based metrics (e.g., ECE) at the dataset level, as it can lead to misperception on the actual calibration quality of a model, and hence, the reliability of its uncertainty estimations. Noticeable is also the agreement among the uncertainty methods at subject-level, suggesting only little benefit in selecting one uncertainty method over another. 

In Table~\ref{tab:ece}, we report for BraTS and the ISIC dataset the following metrics: average subject-level ECE, uncertainty-error overlap (U-E), proportion of correction-benefiting test subjects (BnF), and  Dice coefficient. For a fair comparison, we selected the best-performing threshold for each method whenever the metric required an uncertainty threshold (i.e., U-E and BnF). Overall for both datasets, no uncertainty method outperforms and stands out over the others. Particularly, the \textit{aleatoric} method and methods with large dropout (\textit{center}/\textit{+MC}) yield worst performance. The \textit{aleatoric} method fails to produce uncertainty at the locations of segmentation errors (i.e., low U-E) and is therefore unable to improve segmentation results through corrections, whereas the large dropout mainly negatively affects segmentation performance and ECE. The results further show that MC dropout (\textit{baseline+MC} and \textit{center+MC}) typically improves ECE, U-E, and Dice coefficient over the non-MC versions (\textit{baseline} and \textit{center}), but larger amounts of dropout (\textit{baseline}$<$\textit{center} and \textit{baseline+MC}$<$\textit{center+MC}) results in worse performances, which suggests using MC dropout in the regimes where the benefit with respect to the uncertainty is minimal compared to standard softmax. We could confirm this finding through intermediate dropout strategies (not shown). We also observe good performances of the auxiliary networks, which are typically well-calibrated and profit from a good segmentation performance of their segmentation network (i.e., \textit{baseline} model). In regards to the metrics, we note that low ECE values stem from large amount of low-confident background areas that positively affects the ECE. This also explains the lower ECE values for the BraTS dataset, which contains more background (even with applied brain mask) than the ISIC dataset, due to the additional image dimension. Additionally, the BnF only considers TPU and FPU uncertainties and is therefore favorable for methods with low precision (more FP typically yields more FPU). We found this to be the reason for the bad correction performance of the ensemble on the BraTS dataset, even though the uncertainty-error overlap was good.

\begin{table}[t]
\caption{Performances of the different uncertainties with respect to expected calibration error (ECE), uncertainty-error overlap (U-E), proportion of correction-benefiting test subjects (BnF), and Dice coefficient. Values are presented as \textit{mean (rank)}. Standard deviation is omitted due to marginal differences. Upwards and downwards arrow indicate desired higher and lower metric values, respectively. Horizontal separation group types of uncertainty methods.}
\scriptsize
\centering
\begin{tabular}{l@{\hspace{0.2em}}c@{\hspace{0.5em}}c@{\hspace{0.5em}}c@{\hspace{0.5em}}c@{\hspace{1em}}c@{\hspace{0.5em}}c@{\hspace{0.5em}}c@{\hspace{0.5em}}c}
\toprule
{} & \multicolumn{4}{c}{\textbf{BraTS}} & \multicolumn{4}{c}{\textbf{ISIC}} \\
\cmidrule(lr{1.5em}){2-5} \cmidrule(lr){6-9}
{} &     ECE \% $\downarrow$ &   U-E $\uparrow$ & BnF $\uparrow$ &    Dice $\uparrow$ &     ECE \% $\downarrow$ &   U-E $\uparrow$ & BnF  $\uparrow$&    Dice $\uparrow$\\
\midrule
baseline        &  0.925 (4) &  0.432 (2) &  0.39 (3) &  0.874 (2) &  7.256 (4) &  0.424 (4) &  0.26 (4) &  0.814 (3) \\
center          &  1.758 (7) &  0.409 (5) &   \textbf{0.5} (1) &  0.866 (5) &  9.415 (8) &  0.411 (6) &  0.27 (3) &   0.78 (6) \\[2ex]
baseline+MC     &    \textbf{0.9} (1) &  \textbf{0.433} (1) &  0.36 (4) &  0.874 (2) &   7.36 (5) &  0.428 (3) &  0.24 (5) &  0.813 (4) \\
center+MC       &  1.233 (6) &  \textbf{0.433} (1) &  0.27 (6) &  0.868 (4) &  8.766 (7) &  0.428 (3) &  0.17 (6) &  0.794 (5) \\[2ex]
ensemble        &  0.919 (2) &  \textbf{0.433} (1) &  0.32 (5) &  \textbf{0.879} (1) &  \textbf{7.131} (1) &  0.431 (2) &  0.31 (2) &  \textbf{0.831} (1) \\[2ex]
auxiliary feat. &  0.923 (3) &  0.427 (3) &  0.48 (2) &  0.874 (2) &  7.216 (3) &  0.421 (5) &  \textbf{0.33} (1) &  0.814 (3) \\
auxiliary segm. &  0.925 (4) &  0.412 (4) &  0.48 (2) &  0.874 (2) &  7.212 (2) &  \textbf{0.433} (1) &  0.27 (3) &  0.814 (3) \\[2ex]
aleatoric       &  1.134 (5) &  0.054 (6) &  0.06 (7) &  0.872 (3) &  7.837 (6) &  0.058 (7) &  0.12 (7) &   0.82 (2) \\
\bottomrule
\end{tabular}
\label{tab:ece}
\end{table}

\section{Discussion}
The results show that although current voxel-wise uncertainty measures are rather well-calibrated at the dataset level (i.e., all voxels in the dataset) they tend to fail at the subject level (Fig. \ref{fig:calibration}). This observation is to be expected since subject-level calibration errors (under- or overcalibration) can average out at the dataset level. Based on the proposed calibration-based metric, no overall best uncertainty measure was found among the studied methods. From our experiments we can conclude that methods that aggregate voxel-wise uncertainty to provide subject-level estimations are not reliable enough to be used as a mechanism to detect failed segmentations. We thus conclude on the importance of developing subject-level uncertainty estimation in medical image segmentation that can cope with the issue of High-Dimension-Low-Sample-Size (HDLSS) to ensure their reliability in practice.

Unsurprisingly, the ensemble method yields rank-wise the most reliable results (Tab.~\ref{tab:ece}) and would typically be a good choice (if the resources allow it). The results also revealed that methods based on MC dropout are heavily dependent on the influence of dropout on the segmentation performance. In contrast, auxiliary networks turned out to be a promising alternative to existing uncertainty measures. They perform comparable to other methods but have the benefit of being applicable to any high-performing segmentation network not optimized to predict reliable uncertainty estimates. No significant differences were found between using \textit{auxiliary feat.} and \textit{auxiliary segm.}. Through a sensitivity analysis performed over all studied uncertainty methods (not shown), we could confirm our observations that different uncertainty estimation methods yield different levels of precision and recall. Furthermore, we observed that when using current uncertainty methods for correcting segmentations, a maximum benefit can be attained when preferring a combination of low precision segmentation models and uncertainty-based false positive removal. 

Our evaluation has several limitations worth mentioning. First, although the experiments were performed on two typical and distinctive datasets, they feature large structures to segment. The findings reported herein may differ for other datasets, especially if these consists of very small structures to be segmented. Second, the assessment of the uncertainty is influenced by the segmentation performance. Even though we succeeded in building similarly performing models, their differences cannot be fully decoupled and neglected when analyzing the uncertainty. 

Overall, we aim with these results to point to the existing challenges for a reliable utilization of voxel-wise uncertainties in medical image segmentation, and foster the development of subject/patient-level uncertainty estimation approaches under the condition of HDLSS. We recommend that utilization of uncertainty methods ideally need to be coupled with an assessment of model calibration at the subject/patient-level. Proposed conditions, along with the threshold-free ECE metric can be adopted to test whether uncertainty estimations can be of benefit for a given task.

\subsubsection*{Acknowledgments.} This work was supported by the Swiss National Foundation by grant number 169607. The authors thank Fabian Balsiger for the valuable discussions.

%
%
%
\bibliographystyle{splncs04}
\bibliography{references_short}
\end{document}